\documentclass[12pt]{iopart}
\usepackage{epsfig}
\newcommand{\spur}[1]{\not\! #1 \,}
\begin{document}
\begin{flushright}
DPT/00/78\\IPPP/00/12
\end{flushright}
\title[Lifetimes of $b$-flavoured hadrons]{Lifetimes of $b$-flavoured
hadrons}
\author{Fulvia De Fazio\dag}
\address{\dag\ Institute for Particle Physics Phenomenology, University
of Durham, DH1 3LE, UK}

\begin{abstract}
I discuss the  heavy quark expansion for the inclusive widths of heavy-light
hadrons, which predicts quite well the experimental ratios
of $B_q$ meson lifetimes. As for
$\Lambda_b$, current determinations of
${\cal O}(m_b^{-3})$ contribution  to $\tau(\Lambda_b)$ do not allow to
explain  the small measured value of $\tau(\Lambda_b)/\tau(B_d)$.
As a final topic, I discuss the  implications of the
measurement of the $B_c$ lifetime.
\end{abstract}
\section{Lifetimes of heavy-light hadrons}

Inclusive particle widths describe the decay of the particle into
all possible final states with given quantum numbers $f$. 
For weakly decaying heavy-light  $Q{\bar q}$ ($Q q q$) hadrons $H_Q$,
the {\it spectator model}
considers  only the heavy quark $Q$ as active in the 
decay, the light degrees of freedom 
remaining unaffected.  Hence, 
all the hadrons containing the same heavy quark $Q$ should have the 
same lifetime;
this  picture should become accurate 
in the  $m_Q \to \infty$ limit, when the heavy quark
decouples from the light degrees of freedom. 
However, the measurement of beauty hadron 
lifetime ratios \cite{blg}:
\begin{equation}
{\tau(B^-) \over \tau(B_d)}=1.066 \pm 0.02 \;\;, \;\;
{\tau(B_s) \over \tau(B_d)}=0.99 \pm 0.05 \;\;, \;\;
{\tau(\Lambda_b) \over \tau(B_d)}=0.794 \pm 0.053 \;\;
\label{ratiolam}
\end{equation}
shows  that
$\tau(\Lambda_b) / \tau(B_d)$ significantly differs from the 
spectator model prediction.

A more refined approach  consists in
computing inclusive  decay widths of $H_Q$ hadrons
as an expansion in powers of $m_Q^{-1}$ \cite{misha}. 
Invoking the optical theorem, one can write
$\Gamma(H_Q \to X_f)=2 Im\langle H_Q |{\hat T}|H_Q\rangle / 2
M_{H_Q}$, with 
${\hat T}=i  \int d^4x  T[{\cal L}_w(x){\cal L}_w^{\dag}(0)]$
the transition operator describing the heavy quark
$Q$ with the same momentum in the 
initial and final state, and ${\cal L}_w$ the effective  lagrangian governing
the decay $Q \to X_f$. 
An operator product expansion of $\hat T$ 
in the inverse mass of the heavy 
quark allows to write:
${\hat T}=\sum_i C_i {\cal O}_i$,
with  the local operators ${\cal O}_i$  ordered 
by increasing dimension, and the
 coefficients $C_i$ proportional to increasing powers of $m_Q^{-1}$.
As a result, for a beauty hadron $H_b$ the general expression of the
width $\Gamma(H_b \to X_f)$ is:
\begin{equation} 
\fl \Gamma(H_b \to X_f)=\Gamma_0    
\Big[c_3^f \langle {\bar b}b \rangle_{H_b} 
+{c_5^f \over m_b^2} 
\langle {\bar b} i g_s \sigma \cdot G b \rangle_{H_b}
+\sum_i {c_6^{f(i)} \over m_b^3} 
\langle {\cal O}_i^6 \rangle_{H_b}  +{\cal O}\Big({1 \over m_b^4}
\Big) \Big]   \;\; ,
\label{ris}
\end{equation}
with $\displaystyle{\langle O
\rangle_{H_b}={\langle H_b|O|H_b\rangle \over 2 M_{H_b}}}$, 
$\displaystyle{\Gamma_0={G_F^2 m_b^5 \over 192 \pi^3}|V_{qb}|^2}$ and 
$V_{qb}$ the relevant CKM  matrix element.

The first operator in (\ref{ris}) is ${\bar b}b$, with dimension 
$D=3$; the chromomagnetic operator ${\cal O}_G={\bar b}{g \over 2} 
\sigma_{\mu \nu} G^{\mu \nu}b$, responsible of the 
heavy quark-spin symmetry breaking, has $D=5$; 
the operators $O_i^6$ have $D=6$.
In the limit $m_b \to \infty$, the heavy quark 
equation of motion allows to write:
\begin{equation}
\langle {\bar b}b\rangle_{H_b}= 1+{\langle {\cal O}_G \rangle_{H_b} \over
2 m_b^2} 
-{ \langle{\cal O}_\pi \rangle_{H_b} \over 2 m_b^2} +{\cal O}\Big({1 \over
m_b^3}\Big) ,
\label{bbarb}
\end{equation}
with ${\cal O}_\pi={\bar b}(i {\vec D})^2b$ the
heavy quark kinetic energy operator.
When combined with (\ref{bbarb}), the first term in (\ref{ris})
reproduces the spectator model result.  
${\cal O}(m_b^{-1})$ terms  are absent \cite{cgg,buv} 
since $D=4$ operators are reducible to  ${\bar b}b$ by the
equation of motion. Finally, the operators
${\cal O}_G$ and ${\cal O}_\pi$ are spectator blind, not sensitive 
to light flavour. Their matrix elements can be determined from experimental
data; as a matter of fact, defining
$\mu_G^2(H_b)=\langle {\cal O}_G \rangle_{H_b}$ and 
$\mu_\pi^2(H_b)=\langle {\cal O}_\pi \rangle_{H_b}$, one has:
$\mu_G^2(B)=3(M_{B^*}^2-M_B^2)/4$, while  
$\mu_G^2(\Lambda_b)=0$
since the light degrees of freedom in the $\Lambda_b$
have zero total angular 
momentum relative to the heavy quark. Moreover, from the mass formula:
$\displaystyle{M_{H_b}=m_b+{\bar \Lambda}+{\mu_\pi^2 -\mu_G^2 \over 2 m_b}
+{\cal O}({m_b^{-2}})}$, with
${\bar \Lambda}$, $\mu_\pi^2$ and  $\mu_G^2$
independent of $m_b$, and from the experimental data, one can infer
 $\mu_\pi^2(B_d)\simeq 
\mu^2_\pi(\Lambda_b)$, as confirmed by QCD 
sum rule  estimates \cite{braun}.

The ${\cal O}(m_b^{-3})$ terms in (\ref{ris})
come from four-quark operators,
 accounting for the presence of the spectator quark in the decay. Their 
general expression is \cite{neub}:
\begin{eqnarray}
O^q_{V-A}=({\bar b}_L \gamma_\mu q_L) ({\bar q}_L \gamma_\mu b_L) \hskip 1
cm &&
T^q_{V-A}=({\bar b}_L \gamma_\mu t^a q_L) ({\bar q}_L \gamma_\mu t^a b_L) 
\nonumber \\
O^q_{S-P}=({\bar b}_R q_L) ({\bar q}_L b_R) \hskip 1 cm &&
T^q_{S-P}=({\bar b}_R t^a q_L) ({\bar q}_L t^a b_R) \label{t} \;.
\end{eqnarray}
 \noindent 
Their matrix elements over  $B_q$ can be parametrized as:
\begin{equation}
\fl \langle O^q_{V-A} \rangle_{B_q} =
\langle O^q_{S-P} \rangle_{B_q}
\Big({m_b+m_q \over M_{B_q}} \Big)=f^2_{B_q} {M_{B_q} \over 8} , \;\;\; 
\langle T^q_{V-A} \rangle_{B_q}= \langle T^q_{S-P} \rangle_{B_q} = 0 \;\;,
\label{fact} 
\end{equation}
$f_{B_q}$ being the $B_q$ decay constant. 
As for $\Lambda_b$, 
one can write:
\begin{equation}
\langle {\tilde O}_{V-A}^q \rangle_{\Lambda_b} =
f^2_B M_B  \;r/48 , 
\hskip 0.5cm
\langle O_{V-A}^q \rangle_{\Lambda_b} =-{\tilde B} 
 \langle {\tilde O}_{V-A}^q\rangle_{\Lambda_b} 
\label{otilde}
\end{equation}
with
 ${\tilde O}_{V-A}^q=({\bar b}_L \gamma_\mu b_L) ({\bar q}_L \gamma_\mu q_L)$.
In the valence quark approximation ${\tilde B}=1$.

Actually, with the computed values of the Wilson coefficients in  (\ref{ris}),
only large values of the parameter $r$ in 
(\ref{otilde}) (namely $r\simeq 3-4$) could explain the observed 
difference between 
$\tau(\Lambda_b)$ and $\tau(B_d)$. This, however, seems not to be the case.

\section{$\langle {\tilde O}^q_{V-A} \rangle_{\Lambda_b}$ from QCD sum rules}

The parameter $r$ in (\ref{otilde}) can be determined using quark models
or lattice QCD \cite{vari}. HQET QCD sum 
rules  allow to estimate it from the correlator:
\begin{equation}
\fl \Pi_{CD}=(1+{\spur v})_{CD} \Pi
(\omega,\omega^\prime) 
= i^2 \int dx  dy \;e^{i \omega v \cdot x-i \omega^\prime v \cdot y} 
\langle 0|T[J_C(x) {\tilde O}_{V-A}^q(0) J_D(y)]|0 \rangle    
\end{equation}
between $\Lambda_b$ interpolating fields $J_{C,D}$ ($C$, $D$ Dirac indices)
\cite{shur}
and the operator ${\tilde O}_{V-A}^q$;
$\omega$ $(\omega^\prime)$ is related to the residual momentum of the incoming 
(outgoing)  current $p^\mu=m_b v^\mu+k^\mu$ with $k^\mu=\omega v^\mu$.
The projection of the interpolating fields on the $\Lambda_b$ state
is parametrized by
$\langle 0|J_C|\Lambda_b(v) \rangle
=f_{\Lambda_b} (\psi_v)_C $ (with $\psi_v$ the spinor for a $\Lambda_b$ of 
velocity $v$).

\begin{figure}
\begin{center}
\mbox{\epsfig{file=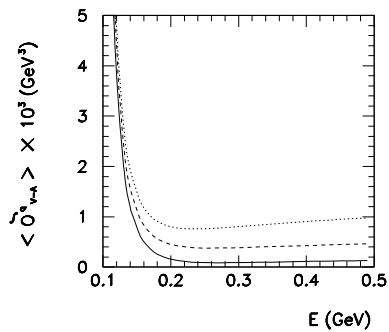,height=1.7in}}
\end{center}
\vskip -0.5cm
\caption{Sum rule for $\langle {\tilde O}_{V-A}^q\rangle_{\Lambda_b}$ as a
function 
of the Borel variable $E$. \label{th271_fig1}}
\end{figure}

Saturating the correlator $\Pi(\omega,\omega^\prime)$  with baryonic
states 
and considering 
the low-lying double-pole contribution 
in the variables $\omega$ and  $\omega^\prime$, one has:
\begin{equation}
\Pi^{had}(\omega, \omega^\prime) = 
\langle{\tilde {\cal O}^q_{V-A}}\rangle_{\Lambda_b}
{f^2_{\Lambda_b} \over 2} \times 
{1 \over (\Delta_{\Lambda_b} - \omega)(\Delta_{\Lambda_b} - \omega^\prime)} 
+ \dots  \label{phad}
\end{equation}
with $\Delta_{\Lambda_b}$  defined by 
$M_{\Lambda_b}= m_b + \Delta_{\Lambda_b}$.
Besides, for negative values of $\omega$, 
$\omega^\prime$, $\Pi$ can be computed in QCD 
in terms of a perturbative contribution and of vacuum condensates:
\begin{equation}
\Pi^{QCD}(\omega, \omega^\prime)= \int d\sigma d\sigma^\prime 
{\rho_\Pi(\sigma, \sigma^\prime) \over(  \sigma - \omega ) (\sigma^\prime
- \omega^\prime)} \label{pope}
\end{equation}
\noindent with possible subtractions omitted \cite{noi}.
The sum rule consists in equating $\Pi^{had}$ 
and $\Pi^{QCD}$.
Moreover, invoking  global duality,
the contribution of  higher resonances and of  continuum to
$\Pi^{had}$  can be  modeled as the QCD term
in the region 
$\omega \ge \omega_c$,
$\omega^\prime \ge \omega_c$, with $\omega_c$ an effective threshold.
Finally,  a double Borel transform  
to  $\Pi^{QCD}$ and $\Pi^{had}$  in  $\omega,
\omega^\prime$, with  Borel parameter $E_1$, $E_2$,
 removes the subtraction terms in (\ref{pope}),
improves factorially  
the convergence of the OPE and enhances
the contribution of the low-lying resonances in $\Pi^{had}$.
Choosing $E_1=E_2=2 E$, one gets a sum rule the result of which is
depicted in
figure \ref{th271_fig1}.
Considering 
the variation with $E$ and the threshold $\omega_c$, one has an estimate of
$\langle\tilde {\cal O}^q_{V-A}\rangle_{\Lambda_b}$:
\begin{equation}
\langle\tilde {\cal O}^q_{V-A}\rangle_{\Lambda_b} \simeq (0.4 - 1.20) \times 
10^{-3} \; GeV^3 \;\;\; , \label{res}
\end{equation}
corresponding to  $r\simeq 0.1-0.3$ \cite{noi}.
The same calculation gives
$\tilde B \simeq 1$. 
This result produces  $\tau(\Lambda_b)/\tau(B_d)\ge 0.94$, at odds with
the experimental result. The discrepancy discloses 
exciting perspectives both from  experimental and theoretical sides
\cite{voloshin}.

\section{$B_c$ lifetime}

A different hadronic system, whose lifetime can be determined by OPE-based
methods, is the $B_c$ meson, observed at Fermilab with mass 
$M_{B_c}=6.40 \pm 0.39\pm 0.13 ~GeV$ and lifetime
$\tau_{B_c}=0.46 \pm^{0.18}_{0.16}\pm0.03~ps$ \cite{cdf}.
Like quarkonium states, $B_c$ can be
treated in a  non relativistic way, but unlike them it can decay only
weakly, with the main decay mechanisms induced by the quark transitions
$b \to c W^- $, ${\bar c}  \to {\bar s} W^- $ and
${\bar c} b \to  W^- $ (annihilation).
Predictions for $\tau_{B_c}$  spread in the range
$0.4-1.2~ps$ \cite{tutti,quigg,beneke}. 
In the $m_b, m_c \to \infty $ limit one would have
$\Gamma_{B_c}=\Gamma_{b,spec}+\Gamma_{c,spec}$. Corrections to this
result can be computed using  an OPE organized in powers of the heavy quark
velocity \cite{beneke}.
The result  is: $\tau_{B_c} \simeq 0.4-0.7~ps$,
together with the prediction of the
dominance of charm transitions;  as a matter of fact, 
$b$-decay dominance would imply a larger lifetime: $\tau_{B_c}=1.1-1.2~ps$
\cite{quigg}. Hence, the measurement of  $\tau_{B_c}$  provides us
with the first hints on the underlying dynamics in this meson.
For this system, 
it is interesting to investigate the validity of the non relativistic
approximation: actually, one estimates 
$\langle {k^2}\rangle / m_c^2\simeq 0.43$, where
$\langle{k^2}\rangle$ is the average squared momentum of the charm quark,
implying possible deviations from the non relativistic limit \cite{noi1}.
\section{Conclusions}
$1/ m_Q$ expansion can be used to compute 
inclusive widths of heavy-light hadrons. A  QCD sum rule calculation 
of the matrix element $\langle {\tilde O}^q_{V-A} \rangle_{\Lambda_b}$ 
contributing to ${\cal O}(m_b^{-3})$ to the $\Lambda_b$ lifetime
gives the result: $\tau(\Lambda_b)/\tau(B_d)\ge 0.94$, thus implying
that such a correction  does not explain the observed difference
between $\tau(\Lambda_b)$ and $\tau(B_d)$. 
Finally,  the measurement of $B_c$
lifetime already enlightens some aspects of the quark dynamics in this
meson. 

\ack
I thank P. Colangelo for  collaboration on the topics discussed above.
I acknowledge for support  the EU-TMR Programme, Contract
No. CT98-0169, EuroDA$\Phi$NE.


\section*{References}
\begin{thebibliography}{99}
\bibitem{blg} LEP $B$ lifetime group,
              http://wwwcn.cern.ch/~claires/lepblife.html.
\bibitem{misha} I. Bigi {\it et al.}, in S. Stone (ed.), $B$ Decays, 2nd
ed., World Scientific, pag.132.
\bibitem{cgg} J. Chay {\etal.}, {\it Phys. Lett.} {\bf B 247} (1990)   
  399.
\bibitem{buv} I. Bigi {\etal.}, {\it Phys. Lett.} {\bf B 293} (1992)
              430; {\bf 297} (1993) 477 (E).
\bibitem{neub} M. Neubert and C.T. Sachrajda, {\it Nucl. Phys.} {\bf B
483} (1997) 339.
\bibitem{braun} P. Ball {\etal.}, {\it Phys. Rev.} {\bf D 49}
(1994) 2472; P. Colangelo {\etal.}, {\it Phys.  Rev.} {\bf D 54}
(1996) 4622.
\bibitem{vari} B. Guberina {\etal.}, {\it Z. Phys.} {\bf  C 33}
 (1986) 297; J.L. Rosner, {\it Phys. Lett.} {\bf B 379} (1996) 267;
 M. Di Pierro {\it et al.}, UKQCD Collab. hep-lat/9906031.
\bibitem{shur} E.V. Shuryak, {\it Nucl. Phys.} {\bf B 198} (1982) 83.
\bibitem{noi} P. Colangelo and F. De Fazio, {\it Phys. Lett.} {\bf B
387} (1996) 371.
\bibitem{voloshin}
M.~B.~Voloshin,
hep-ph/0011099.
\bibitem{cdf}CDF Collab., {\it Phys. Rev. Lett.} {\bf 81}
(1998) 2432; {\it Phys. Rev.} {\bf D58} (1998) 112004.
\bibitem{tutti}M. Lusignoli and M. Masetti, {\it Z. Phys.} {\bf C 51}
(1991) 549; P. Colangelo {\etal.}, {\it Z. Phys.  } {\bf C 57}
(1993) 43; A.Y. Anisimov {\etal.}, {\it Phys. Lett. } {\bf B452}
(1999) 129.
\bibitem{quigg}C. Quigg, in {\it Proceedings of the Workshop on $B$
Physics at Hadron Accelerators}, Snowmass, Colorado 1993, ed. P. McBride
and C.S. Mishra, pag. 439.
 \bibitem{beneke} M. Beneke and G. Buchalla, {\it
Phys. Rev.} {\bf D53} (1996)4991.
\bibitem{noi1}P. Colangelo and F. De Fazio, {\it Mod. Phys. Lett.} {\bf
A14} (1999) 2303; {\it Phys. Rev.} {\bf D61}
(2000) 034012.


\end{thebibliography}
\end{document}